\documentclass[reprint,superscriptaddress,amsmath,amssymb,aps,prb]{revtex4-1}
\usepackage{graphicx}
\usepackage{placeins}
\usepackage{bm}
\usepackage{xcolor}
\usepackage{soul}

\begin{document}

\title{Spiraling vortices in exciton-polariton condensates}

\author{Xuekai Ma}
\email{xuekai.ma@gmail.com}
 \affiliation{Department of Physics and Center for Optoelectronics and Photonics Paderborn (CeOPP), Universit\"{a}t Paderborn, Warburger Strasse 100, 33098 Paderborn, Germany}

\author{Yaroslav V. Kartashov}%
\affiliation{ 
Institute of Spectroscopy, Russian Academy of Sciences, Troitsk, Moscow, 108840, Russia
}%
\affiliation{ 
ICFO-Institut de Ci{\`e}ncies Fot{\`o}niques, The Barcelona Institute of Science and Technology, 08860 Castelldefels (Barcelona), Spain
}%

\author{Tingge Gao}
\affiliation{%
Tianjin Key Laboratory of Molecular Optoelectronic Science, Institute of Molecular Plus, Tianjin University, Tianjin 300072, China
}%
\affiliation{%
Department of Physics, School of Science, Tianjin University, Tianjin 300072, China
}%

\author{Lluis Torner}%
\affiliation{ 
ICFO-Institut de Ci{\`e}ncies Fot{\`o}niques, The Barcelona Institute of Science and Technology, 08860 Castelldefels (Barcelona), Spain
}%

\author{Stefan Schumacher}
 \affiliation{Department of Physics and Center for Optoelectronics and Photonics Paderborn (CeOPP), Universit\"{a}t Paderborn, Warburger Strasse 100, 33098 Paderborn, Germany}
 \affiliation{College of Optical Sciences, University of Arizona, Tucson, AZ 85721, USA}
 
\date{\today}

\begin{abstract}
We introduce the phenomenon of spiraling vortices in driven-dissipative (non-equilibrium) exciton-polariton condensates excited by a non-resonant pump beam. At suitable low pump intensities, these vortices are shown to spiral along circular trajectories whose diameter is inversely proportional to the effective mass of the polaritons, while the rotation period is mass independent. Both diameter and rotation period are inversely proportional to the pump intensity. Stable spiraling patterns in the form of complexes of multiple mutually-interacting vortices are also found. At elevated pump intensities, which create a stronger homogeneous background, we observe more complex vortex trajectories resembling Spirograph patterns. 
\end{abstract}

\maketitle

\section{Introduction}
Vortices, or wavefront singularities, nested in wave or matter fields evolving in nonlinear media are fascinating topological objects that have been widely studied in various physical systems, including superfluids~\cite{yarmchuk1979observation}, superconductors~\cite{blatter1994vortices}, atomic condensates~\cite{matthews1999vortices,madison2000vortex,fetter2001vortices,fetter2009rotating}, optics~\cite{swartzlander1992optical,desyatnikov2005optical}, and exciton-polaritons~\cite{lagoudakis2008quantized,sanvitto2010persistent,roumpos2011single}. They are characterized by a discrete topological charge, or winding number, given by the  phase accumulated around a closed contour surrounding the phase dislocation where the density/intensity of the wave must vanish. Vortices are ubiquitous also in linear media, but they show specific features in systems that exhibit a nonlinear response. In particular, the latter may compensate diffractive broadening and result in the formation of self-sustained vortex solitons that can be \textit{bright} (i.e., have the form of a localized vortex ring) or \textit{dark} (i.e., a localized low-density feature nested in an extended flat-top wave) in media with attractive or repulsive nonlinearity, respectively. Such vortex solitons in uniform conservative nonlinear media are radially symmetric structures that can be excited by various techniques. In atomic Bose-Einstein condensates the loss of particles may result in the spiraling motion of the vortices towards the edge of the cloud and towards the regions where they vanish~\cite{fedichev1999dissipative,fetter2001vortices,jackson2009finite,fetter2009rotating,rooney2010decay}.

\begin{figure} [!b]
\includegraphics[width=1\columnwidth]{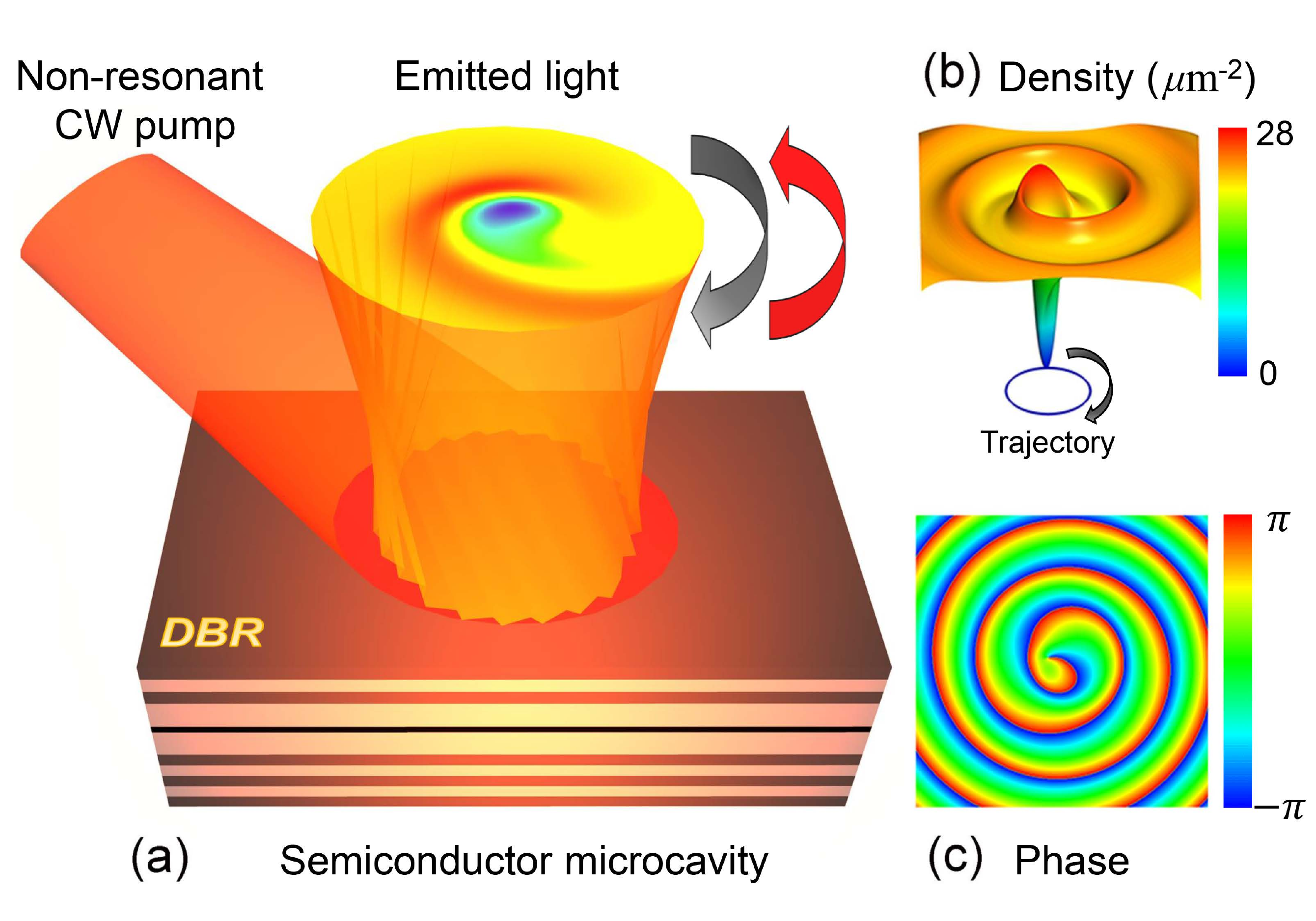}
\caption{(a) Sketch of a planar semiconductor quantum-well microcavity with non-resonant continuous-wave (CW) optical excitation. The light emitted from the surface of the distributed Bragg reflector (DBR) shows the presence of a spiraling vortex. The rotation direction of the phase singularity (condensate) is indicated by the gray (red) arrow. (b) Density profile $|\Psi|^2$ (in $\mu m^{-2}$) of a spiraling vortex structure with its core travelling along a closed trajectory as indicated by the arrow. (c) Phase $\textrm{arg}(\Psi)$ of the spiraling vortex in (b).}\label{sample}
\end{figure}

Exciton-polariton condensates form in the regime of strong coupling between photons and excitons in a semiconductor microcavity, inheriting the properties of their optical and matter constituents. The former  lends a small effective mass to the polaritons, while the latter introduces strong repulsive nonlinearity due to exciton-exciton interactions. Polaritons may undergo condensation~\cite{deng2002condensation,kasprzak2006bose} and can be excited by both resonant and non-resonant pumping. Under resonant excitation, the pump determines the phase profile of the vortex wavefunction~\cite{sanvitto2010persistent,dominici2015vortex,zezyulin2019vortex,dominici2018vortex}. A non-resonant pump with its energy far above the excitonic resonance creates an excitation reservoir that determines the condensate formation dynamics and spatial distribution. Due to the non-equilibrium nature and reservoir-polariton interactions, both bright~\cite{dall2014creation,ma2020realization} and dark vortex states~\cite{lagoudakis2008quantized,liew2015instability} can be excited. 

Generally, a symmetric single polariton vortex does not move in a uniform background~\cite{warszawski2012unpinning}, while in a non-uniform background it can spiral out from the center ~\cite{ostrovskaya2012dissipative} as in dissipative Bose-Einstein condensates, unless a trapping potential is used to prevent the motion to the periphery~\cite{borgh2012robustness}. The interaction between vortices typically leads to creeping, irregular oscillations, or even mutual annihilation~\cite{rosanov2005curvilinear,fraser2009vortex,nardin2011hydrodynamic,liew2015instability,dominici2015vortex}. However, in a non-equilibrium polariton condensate, multiple spontaneously generated vortices may create non-conventional vortical structures showing periodic evolution dynamics~\cite{liew2015instability}, whose origins remain unexplored. Spiraling waves have been studied in the context of the complex Ginzburg-Landau equation in the presence of considerable diffusion~\cite{PhysRevE.62.7627,RevModPhys.74.99}. However, in such works the vortex profiles are spatially symmetric and the core may become unstable in the parameter range where diffusive effects are weak compared to dispersion.

In this paper, we report on the formation of \textit{spiraling vortices} in a planar semiconductor microcavity without a built-in external potential and only using non-resonant optical excitation as sketched in Fig.~\ref{sample}(a). In the resulting self-sustained dynamical state of a spiraling wave, the vortex nested in its center [Fig.~\ref{sample}(b)] tends to drive any other phase singularities away from it to a certain minimal distance. At low pump intensities the phase singularity as such is very stable but persistently moves along a helical trajectory in the $(x,y,t)$ space [Fig.~2(a)]. The phase distribution in such structures is also spiraling in the entire $(x,y)$ plane [Fig.~\ref{sample}(c)], even in the regions far from the vortex center, where the background is nearly homogeneous. Remarkably, the direction of motion of the phase singularity in the transverse plane and the tangential component of the outgoing propagating currents in the condensate are found to be opposite to each other, as indicated by the arrows in [Fig.~\ref{sample}(a)]. Higher pump intensities lead to considerably more complex, but still regular vortex trajectories. We predict that the described dynamics should be readily observable and show that it occurs for a broad range of pump intensities in a physical setting that features experimentally available effective polariton masses and polariton decay rates.

\section{Theoretical model}
The dynamics of a polariton condensate in a semiconductor microcavity at the bottom of the lower-polariton branch is described by the driven-dissipative Gross-Pitaevskii (GP) equation coupled to the density of the exciton reservoir~\cite{wouters2007excitations}:
\begin{equation}\label{e1}
\begin{aligned}
i\hbar\frac{\partial\Psi(\mathbf{r},t)}{\partial t}&=\left[-\frac{\hbar^2}{2m_{\text{eff}}}\nabla_\bot^2-i\hbar\frac{\gamma_\text{c}}{2}+g_\text{c}|\Psi(\mathbf{r},t)|^2 \right.\\
&+\left.\left(g_\text{r}+i\hbar\frac{R}{2}\right)n(\mathbf{r},t)\right]\Psi(\mathbf{r},t), \\
\end{aligned}
\end{equation}
\begin{equation}\label{e2}
\frac{\partial n(\mathbf{r},t)}{\partial t}=\left[-\gamma_r-R|\Psi(\mathbf{r},t)|^2\right]n(\mathbf{r},t)+P(\mathbf{r},t)\,.
\end{equation}
Here $\Psi(\mathbf{r},t)$ is the polariton wavefunction and $n(\mathbf{r},t)$ is the exciton reservoir density. The effective mass of polaritons is $m_{\text{eff}}=a\times10^{-4}m_{\text{e}}$, where $a$ is a variable mass coefficient and $m_{\text{e}}$ is the free electron mass. We include a finite polariton lifetime of $3\,\mathrm{ps}$ with $\gamma_\text{c}=0.33$ ps$^{-1}$ and a reservoir decay of $\gamma_\text{r}=1.5\gamma_\text{c}$. The polariton condensate is replenished in a stimulated manner by the coupling to the reservoir density $n(\mathbf{r},t)$ with $R=0.01$ $\mu$m$^{2}$ ps$^{-1}$. The reservoir is excited by a non-resonant spatially homogeneous pump $P(\mathbf{r},t)=P_0P_{\text{thr}}$, where $P_{\text{thr}}=\gamma_{\text{c}}\gamma_{\text{r}}/R$ is the threshold pump intensity above which condensation occurs, while $P_0>1$ is the dimensionless factor introduced for convenience. Far from the vortex core, in the region where the condensate becomes homogeneous, its density (which does not change with time, in contrast to the density in close proximity of the vortex core) is given by $|\Psi_{\text{hs}}|^2=(P_0-1)\gamma_r/R$. In this region, the exciton reservoir density is given by $n=P/(\gamma_\text{r}+R|\Psi_\text{hs}|^2)=\gamma_\text{c}/R$. These relations follow from Eqs.~(1) and (2), if one requires that losses in the homogeneous region are exactly compensated by the pump and that $\partial n/\partial t=0$ in this region. The interaction strength between polaritons is given by $g_\text{c}=6$ $\mu$eV $\mu$m$^{2}$ and the interaction strength between polariton and reservoir is given by $g_\text{r}=2g_\text{c}$. These are typical parameters for GaAs-based semiconductor microcavities. Substitution of the expression for the asymptotic reservoir density $n=P/(\gamma_\text{r}+R|\Psi_\text{hs}|^2)$ into Eq.~(1) yields a conservative nonlinearity of the form $g_\text{c}|\Psi_\text{hs}|^2+g_\text{r}P(\gamma_\text{r}+R|\Psi_\text{hs}|^2)^{-1}$, which in the low-density limit transforms into $(g_\text{c}-g_\text{r}P_0\gamma_c/\gamma_r)|\Psi_\text{hs}|^2$ (at higher densities one has to take into account contributions from the higher powers of $|\Psi_\text{hs}|^2$), indicating that the reservoir does affect polariton-polariton interactions and, thus, the characteristic scales of the vortices, if they are present in the wavefunction $\Psi$. The spiraling vortices reported here are obtained by numerically solving  Eqs.~(\ref{e1}) and (\ref{e2}) for different system parameters, as indicated below. 

\section{Rotation dynamics}
\begin{figure} 
\includegraphics[width=1.0\columnwidth]{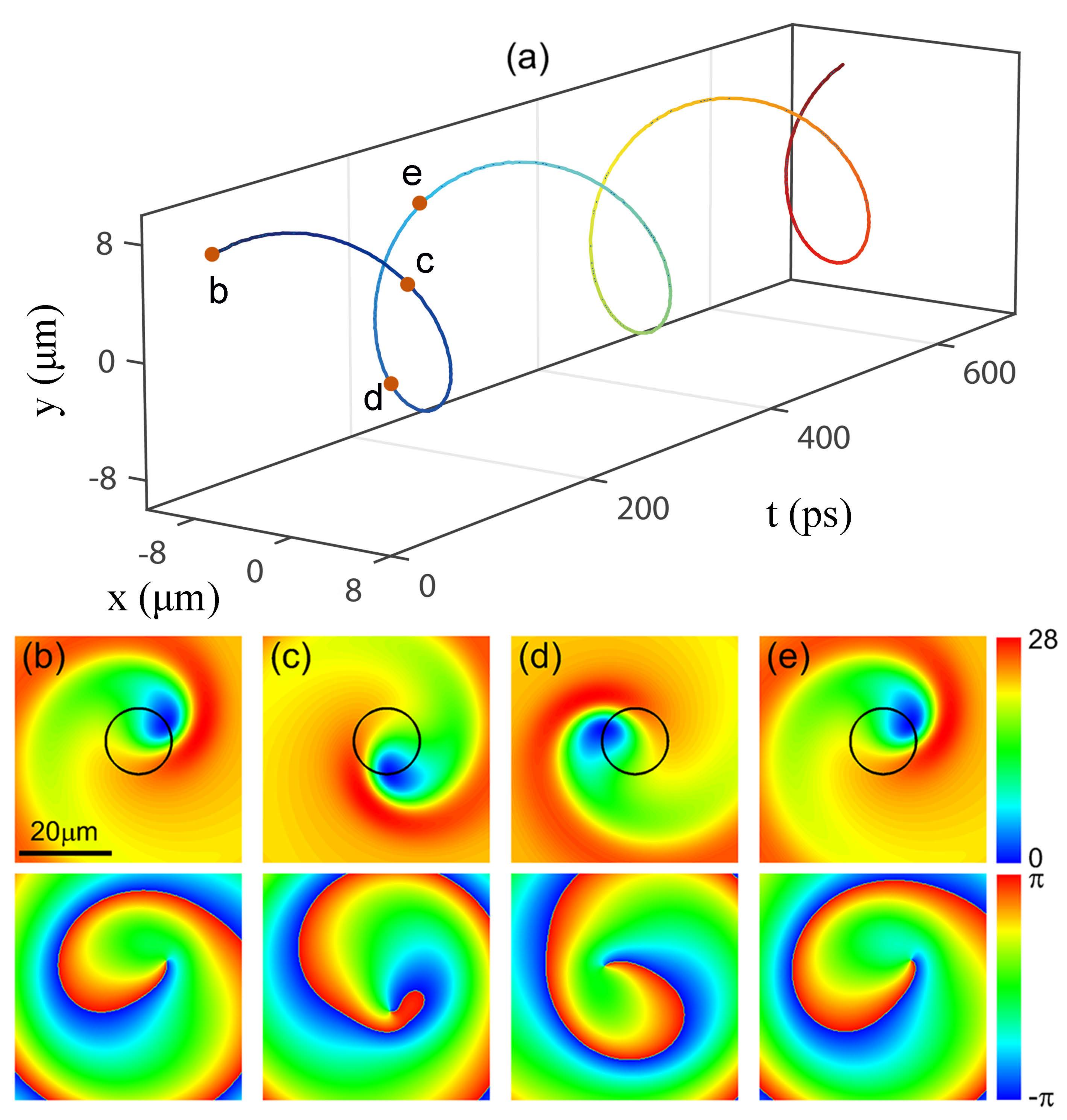}
\caption{(a) Time-dependent trajectory of a phase singularity. (b-e) Density (in $\mu m^{-2}$) (upper row) and phase (lower row) distributions of the spiraling vortex with topological charge $m=1$ in (a) at different moments in time, (b) $t=0$ ps, (c) $t=75$ ps ($\sim P_\textrm{T}/3$), (b) $t=150$ ps ($\sim 2P_\textrm{T}/3$), and (b) $t=225$ ps ($\sim P_\textrm{T}$), corresponding to the brown points in (a). $P_\textrm{T}$ is the temporal helix period. 
}
\label{rotations}
\end{figure}

The helical trajectory of a representative spiraling vortex in the $(x,y,t)$ space is illustrated in Fig.~\ref{rotations}(a), showing periodic vortex rotation around the $x,y=0$ point with time. Figures \ref{rotations}(b)-\ref{rotations}(d) show snapshots of density and phase distributions at different moments in time within one rotation period. Interestingly, one can see from the density profiles that the phase singularity rotates clockwise, while the tangential component of the current direction in the condensate (determined by its phase distribution) is actually counter-clockwise, as indicated by the gray and red arrows in Fig.~\ref{sample}(a). The rotation direction of the condensate is linked to the sign of the topological charge of the spiraling vortex, i.e., a positive (negative) charge corresponds to an overall counter-clockwise (clockwise) condensate rotation. To check the robustness of such behavior, we perturbed the obtained spiraling solutions $\Psi$ by adding complex (amplitude and phase) broadband noise and let them evolve over long times. No instabilities were observed in the numerical evolution. We have also verified that spiraling vortices excited by a broad super-Gaussian pump with a flat-top shape are indistinguishable from those excited by a plane-wave pump. 

\begin{figure} 
\includegraphics[width=1\columnwidth]{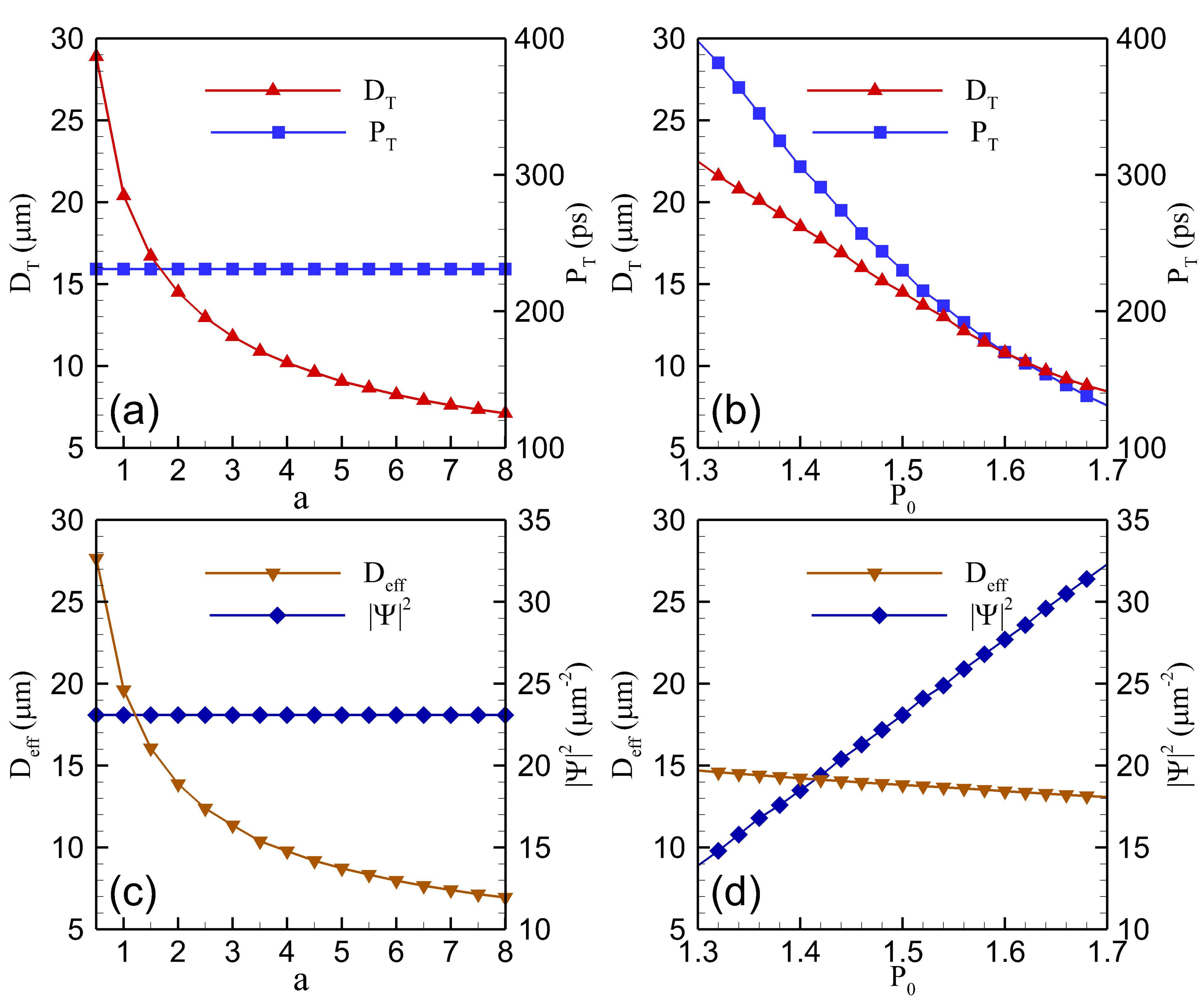}
\caption{Dependence of the diameter ($D_\textrm{T}$) and temporal period ($P_\textrm{T}$) of the helical trajectory of the vortex on (a) the effective mass of polaritons and (b) pump intensity. Dependencies of the diameter of the vortex core $D_\textrm{eff}$ and the background density $|\Psi|^2$ on (c) the effective mass of polaritons and (d) pump intensity. $P_0=1.5$ for (a,c) and $a=2$ for (b,d).}\label{relations}
\end{figure}

The dynamics of the spiraling vortex strongly depend on the effective mass of the condensate. Figure~\ref{relations}(c) shows that the diameter of the vortex core $D_{\text{eff}}$ is inversely proportional to the square-root of the effective mass. Figure~\ref{relations}(a) shows that the diameter of the vortex trajectory $D_\textrm{T}$ has a similar trend ($\sim m_\text{eff}^{-1/2}$) as the effective diameter of the vortex core $D_{\text{eff}}$ from Fig.~\ref{relations}(c). The values of $D_\textrm{T}$ and $D_{\text{eff}}$ are almost the same for $P_0=1.5$, as shown in Figs.~\ref{relations}(a) and \ref{relations}(c). An essential factor that impacts the spiraling vortex is the pump intensity. For low pump intensities only slightly above the condensation threshold the background density $|\Psi_\textrm{hs}|^2 \sim (P_0-1)$ is relatively low, leading to larger sizes of the vortex core $\sim (P_0-1)^{-1}$ and large diameters $D_\textrm{T}$ of the rotation trajectory at fixed effective mass [see Figs.~\ref{relations}(b) and \ref{relations}(d)]. As the pump intensity increases, $D_{\text{T}}$ decreases and leads to larger angular rotation velocities and smaller helix periods, which approximately behave as $P_\textrm{T} \sim D_\textrm{T}^{1/2}$ [Fig. \ref{relations}(b)].

\begin{figure} 
\centering
\includegraphics[width=1\columnwidth]{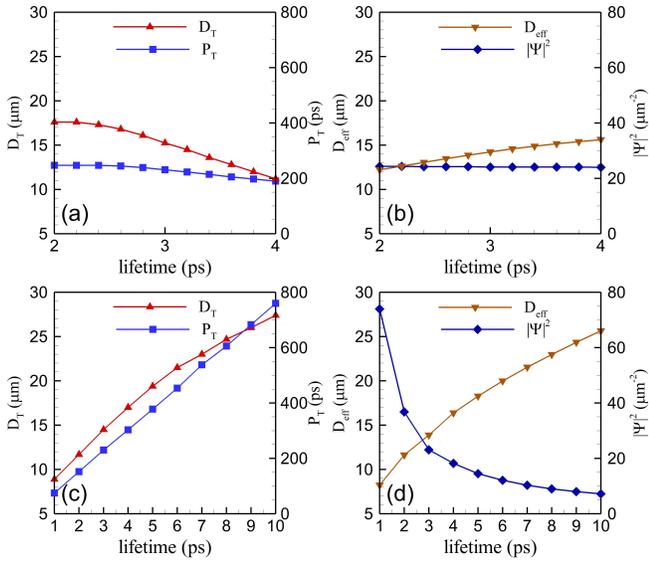}
\caption{Dependence of the diameter ($D_\textrm{T}$) and temporal period ($P_\textrm{T}$) of the helical trajectory of the vortex on polariton lifetime when (a) $P_0$ varies according to $P_0=0.165/\gamma_\text{c}+1$ and (c) when it is fixed at $P_0=1.5$. Dependence of the diameter of the vortex core ($D_\textrm{eff}$) and the background density ($|\Psi|^2$) on polariton lifetime when (b)  $P_0$ varies as $P_0=0.165/\gamma_\text{c}+1$ and (d) for $P_0=1.5$. Here, $a=2$ and $\gamma_\text{r}=1.5\gamma_\text{c}$.}\label{relationSM}
\end{figure}

The dynamics of the spiraling vortices also depend on the lifetime ($1/\gamma_\text{c}$) of polaritons, in addition to the dependence on the effective polariton mass and pump intensity. When studying the dependence of the dynamics of the spiraling vortices on the lifetime of polaritons, we first make sure that the background density of the vortex defined by $|\Psi_{\text{hs}}|^2=(P_0-1)\gamma_r/R$, remains unchanged. This requires proper adjustment of the pump intensity $P_0$ in accordance with the selected $\gamma_\text{c}$ (and, hence, $\gamma_\textrm{r}$) value, since we assume that the ratio $\gamma_\text{r}\equiv1.5\gamma_\text{c}$ is fixed. Having the same background density for different lifetimes, as shown in Fig.~\ref{relationSM}(b), one can observe that the size of the vortex core ($D_\text{eff}$) gradually increases with increase of the polariton lifetime [Fig.~\ref{relationSM}(b)]. The increased spatial coherence of the condensate allows the vortex to experience a stronger influence from the background density, so that for a fixed background density, both the diameter of the spiraling trajectory ($D_\textrm{T}$) and its temporal period ($P_\textrm{T}$) decrease when the lifetime increases [see Fig.~\ref{relationSM}(a)]. In Figs.~\ref{relationSM}(a) and \ref{relationSM}(b), we considered the pump intensity factor to be $P_0=0.165/\gamma_\text{c}+1$, i.e., $P_0=1.5$ for $\gamma_\text{c}=0.33$. 

\begin{figure}
\includegraphics[width=1\columnwidth]{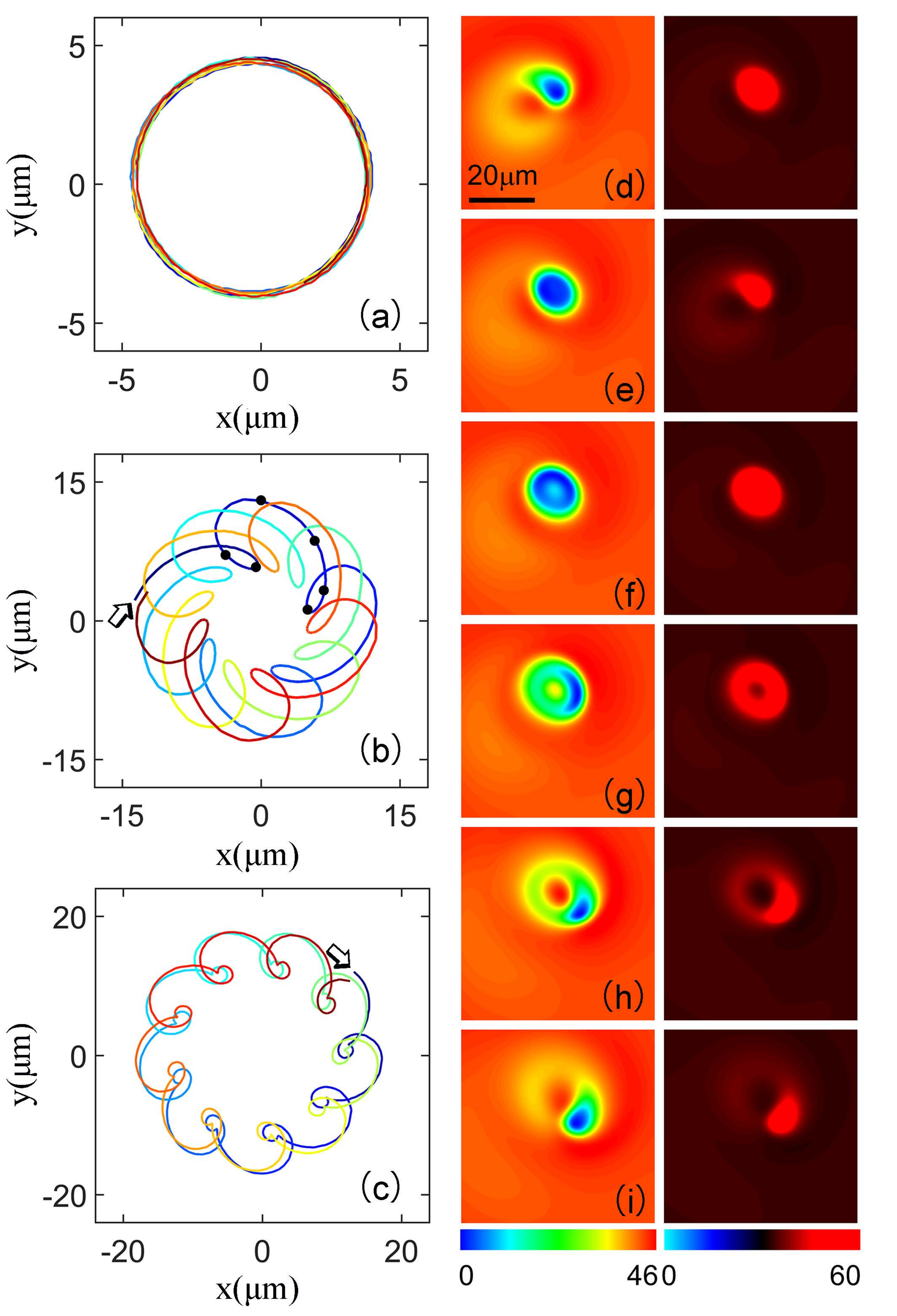}
\caption{(a)-(c) Trajectories of spiraling vortices with $a=2$ for different pump intensities with (a) $P_0=1.7$, (b) $P_0=1.9$, and (c) $P_0=2.1$. Different colors in the trajectory plots are used to distinguish positions at different moments in time of several subsequent rotation periods. The arrows in (b) and (c) indicate the starting points ($t=0$~ps). (d)-(i) Density (in $\mu m^{-2}$) profiles of the condensate (left column) and the reservoir (right column) at $t=$45, 70, 80, 90, 105, and 120~ps, respectively, corresponding to the black points in (b).}\label{patterns}
\end{figure}

For a set ratio of the pump intensity to the condensation threshold, e.g., $P_0=1.5$, when the polariton lifetime increases the reduction of the loss rate results in a decrease of the condensation threshold, $P_\text{thr}=\gamma_\text{c}\gamma_\text{r}/R$. This implies a reduction of the pump strength $P=P_0 P_\textrm{thr}$ with a decrease of $\gamma_\textrm{c}$, which, in turn, causes a reduction of the condensate density, as shown in Fig.~\ref{relationSM}(d). As a consequence, the size of the vortex core, characterized by $D_\text{eff}$, grows due to the weakening of the suppression from the background density. The diameter ($D_\text{T}$) and temporal period ($P_\text{T}$) of the helical trajectory of the spiraling vortex show a similar trend and increase with increase of the polariton lifetime $ 1/\gamma_\textrm{c}$ [Fig.~\ref{relationSM}(c)]. Importantly, the results shown in Figs.~\ref{relationSM}(c) and \ref{relationSM}(d) indicate that the spiraling vortices can exist for a broad range of polariton lifetimes.

\begin{figure} 
\centering
\includegraphics[width=1\columnwidth]{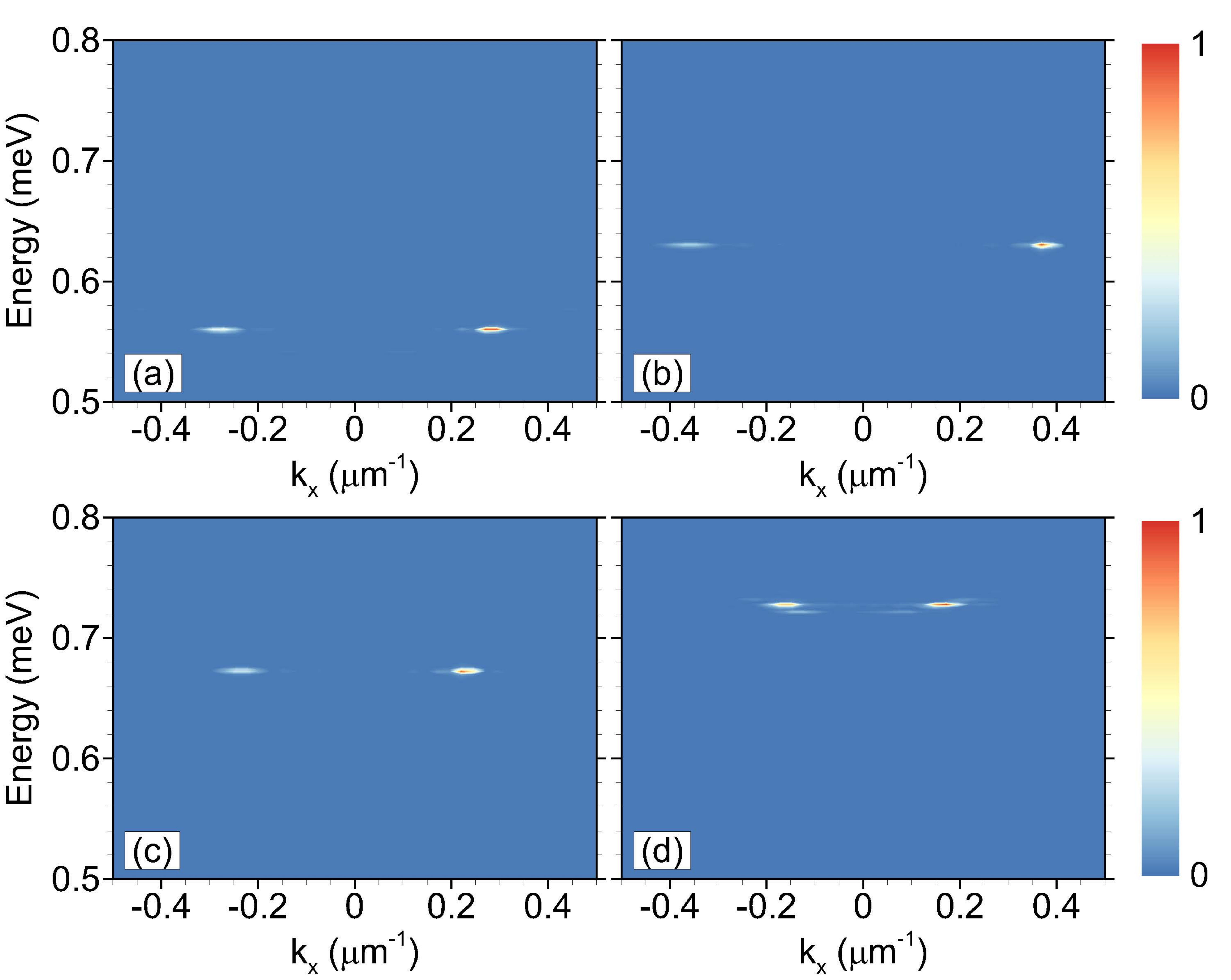}
\caption{Spectra of spiraling vortices at (a) $P_0=1.5$, (b) $P_0=1.7$, (c) $P_0=1.9$, and (d) $P_0=2.1$, corresponding to the solutions in Figs. 2, 4(a), 4(b), and 4(c), respectively, in the main text with $a=2$ and $\gamma_\text{c}=0.33$ (lifetime is 3 ps). The data show the Fourier transform of the time evolution of solution in the $x-t$ plane at $y=0$.}\label{FT}
\end{figure}

Besides slowing down the circular motion of the vortex, a larger pump intensity also increases the density of the reservoir. As a result, the reservoir-condensate interaction becomes stronger and it starts affecting the trajectory of the vortex. It is readily visible that the trajectory in Fig.~\ref{patterns}(a) corresponding to $P_0=1.7$ 
already slightly deviates from the ideal circle, even though the deviation is small. As the pump intensity increases, the vortex trajectory gradually transforms into a Spirograph pattern [Fig.~\ref{patterns}(b)]. In this case, the reservoir density that creates a repulsive potential for polaritons shows a clear peak at the position of the vortex core [Fig.~\ref{patterns}(d)]. During the temporal evolution, the strong reservoir-induced potential repels the condensate and creates an outward flow leading to the broadening of the vortex core and further contraction of the localized spot in the reservoir density [Fig.~\ref{patterns}(e)]. The sharp reservoir density spot then leads to an increase of the condensate density in the middle of the broad vortex core [Fig.~\ref{patterns}(f)] and the simultaneous depletion of the reservoir in the center of the localized spot [Fig.~\ref{patterns}(g)]. A growing condensate density leads to the displacement of the vortex core as shown in Figs.~\ref{patterns}(h-i) and to the subsequent accumulation of reservoir density within the core. These out-of-phase density oscillations lead to a Spirograph trajectory of the phase singularity. Such trajectory can further change for even larger pump intensities [Fig.~\ref{patterns}(c)]. When the homogeneous background becomes even larger as the pump is increased further, the vortex can be spatially pinned. A video showing the temporal evolution of a spiraling state as the pump intensity increases is provided as Supplementary Material~\cite{videos}. 

Since the pump used in the present work is non-resonant, the definition of the excitation energy of the spiraling vortex is not straightforward. However, one can still extract information about the energy/frequency and momentum information from the Fourier transform of their temporal evolution. To that end, we store the evolution of the complex amplitude of the polariton field along the line $y=0$ passing through the center of the rotation trajectory, over several rotation periods, and then Fourier transform the recorded $x-t$ data into the momentum-energy domain, related to the far-field photoluminescence spectra that are measured in the experiments. Due to the unique rotation period and fixed topological charge carried by the vortex, after the Fourier transform well defined energy/frequency and momentum of the vortex can be obtained. Illustrative numerical spectra for different spiraling vortices at the different pump intensities studied are shown in Fig.~\ref{FT}. One can see from Fig.~\ref{FT}(a) that the asymmetric distribution of the spiraling vortex in real space gives rise to an asymmetric spectrum in the $k$-space. When the pump intensity increases for vortices showing circular motion, the reduced diameter ($D_\text{T}$) of the helical trajectory and the vortex core ($D_\text{eff}$) lead to a broader distribution in the momentum space, as shown in Fig.~\ref{FT}(b). When solutions transform into states exhibiting the Spirograph trajectory shown in Figs.~4(b) and 4(c), the width of the distribution in the momentum space starts decreasing with an increase of the pump intensity [see Figs.~\ref{FT}(c) and \ref{FT}(d)]. In most cases spiraling vortices have well-defined energy values. 

\section{Excitation dynamics}
Generally, spiraling vortices can build up from interacting vortices in a polariton condensate~\cite{liew2015instability}. However, this process is spontaneous and uncontrollable. To initialize a spiraling vortex state in a selected region inside the microcavity condensate, which initially does not carry any vorticity, we first excite a vortex that forms in the center of a ring-shaped profile of a non-resonant pump beam~\cite{ma2016incoherent}. Under such conditions, a spiraling phase distribution forms spontaneously but the phase singularity located at the center remains spatially pinned. Subsequently, changing the pump shape from a localized ring to a plane wave typically conserves the central phase singularity and leads to a gradual increase of the condensate density in the entire transverse plane with respect to the density of the homogeneous wave. Simultaneously, the disappearance of the ring-shaped potential created by the ring pump releases the twisted phase carried by the vortex and the vortex core, which rotates supported by the homogeneous background density in the opposite direction compared to the current in the condensate. The balance of the rotation of the vortex core and the current of the condensate results in the excitation of the spiraling vortex. When the pump intensity only slightly exceeds the threshold value, $1.0<P_0<1.3$, the homogeneous background appears to be too weak to host the spiraling vortices. In this case, upon change of the pump shape one observes fast diffraction of the initial vortex. We also observed that a much stronger background density significantly squeezes the size of the vortex core. In some cases the size of the vortex core and the diameter of the trajectory even collapse (see Figs.~\ref{relations}(b) and \ref{relations}(d)), preventing the circular motion of the vortex core.

\section{Complex states}

\begin{figure*} 
\includegraphics[width=1.8\columnwidth]{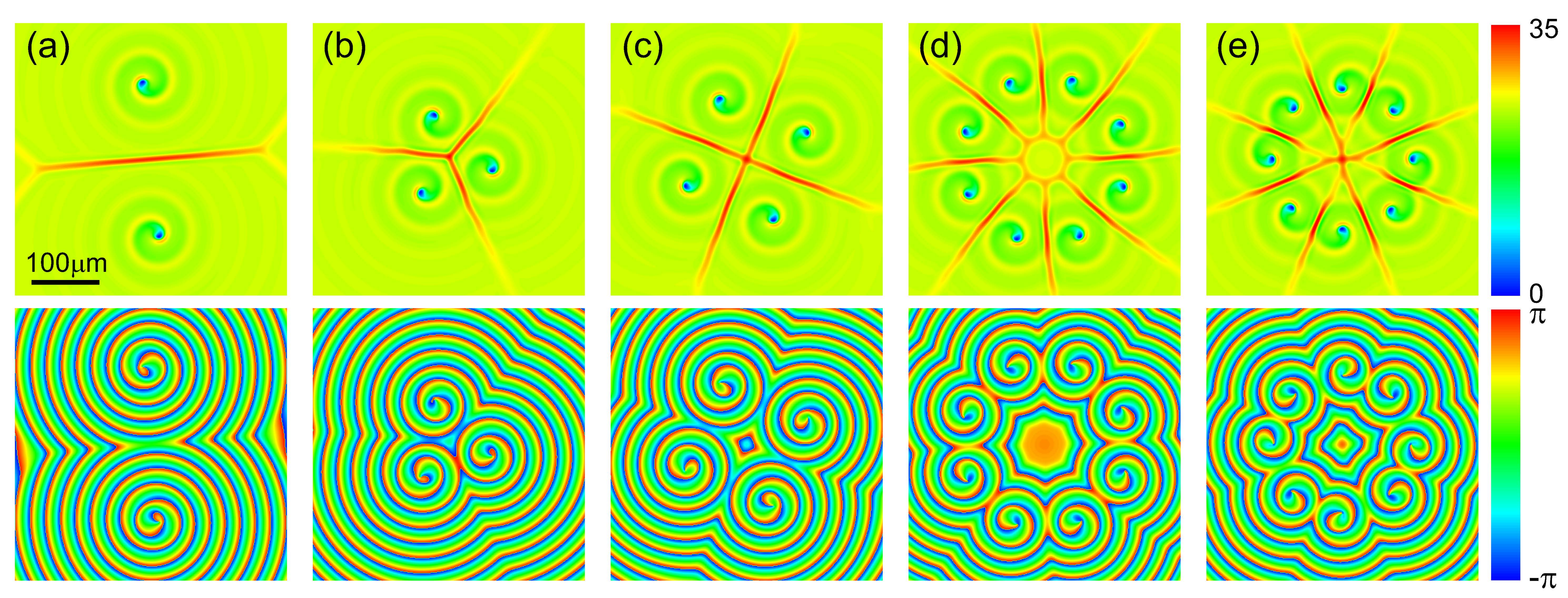}
\caption{Density (in $\mu m^{-2}$) (upper row) and phase (lower row) profiles of spiraling patterns with multiple vortices. The number $N$ (global topological charge $m$ of the pattern) of the spiraling vortices is (a) $N=2$~($m=2$), (b) $N=3$~($m=3$), (c) $N=4$~($m=4$), (d) $N=8$~($m=8$), and (e) $N=8$~($m=0$). Here, $P_0=1.5$ and $a=2$. Videos of the time evolution of the spiraling vortices depicted in (c)-(e) are available in the Supplementary Material~\cite{videos}.}\label{multiple}
\end{figure*}

Previously it was found that increasing the radius of the ring-shaped pump enables the creation of bright vortices with higher charge~\cite{ma2016incoherent}. In the present work, we find that such vortex states with higher charge allow for the creation of complexes of spiraling vortices where the central phase singularity splits into several separate singularities ~\cite{sanvitto2010persistent,kwon2019direct}. Figure~\ref{multiple} shows representative patterns containing different numbers of spiraling vortices. In most cases the number of emerging spiraling vortices is equal to the topological charge of the initial phase singularity, as shown in Figs.~\ref{multiple}(a)-\ref{multiple}(d). A special case is presented in Fig.~\ref{multiple}(e), where four vortex-anti-vortex pairs are formed. This situation arises as a result of the quadrupole-like initial condition with flat phase, i.e. $m=0$, so that the global topological charge of the pattern also remains zero. The dynamics and parameters of each spiraling vortex in the cluster depicted in Fig.~\ref{multiple} are the same as for the single spiraling vortex shown in Fig.~\ref{rotations} with the exception of  the phase and, in the case $m=0$, the direction of rotation (see [\onlinecite{videos}]). Namely, each vortex in the complexes moves along its own spiraling trajectory. 

\section{Conclusions}
We have shown that single and multiple vortices can form robust spiraling structures in polariton condensates. To excite such structures systematically a sudden change of the continuous-wave pump profile from localized ring-shaped to an extended plane wave shape can be used. The phase singularities originally formed in the ring-shaped pump survive and we find them to evolve into spiraling vortices residing on the extended condensate background. The trajectory of a spiraling vortex forms a ring at low pump intensities and a Spirograph pattern when the pump intensity is sufficiently large. We found that the formation of spiraling vortices is a robust phenomenon that occurs for a broad range of effective masses, pump intensities, and polariton decay rates. Also, we elucidated that it occurs for values of parameters that are readily accessible experimentally in exciton-polariton systems in planar semiconductor microcavities. We also tested the impact of small disorder (with characteristic magnitude ~0.1-0.2 meV) on the dynamics of the spiraling states and found that even though the trajectory of the vortex core does not follow a perfect spiral for elevated disorder levels, the vortices do not become spatially pinned and still move in the transverse plane. Naturally, the deviation from a perfect spiraling trajectory increases with increasing level of disorder. From these findings we conclude that the vortex patterns studied in the present work should be observable with currently available technology but would benefit from the use of samples with reduced spatial inhomogeneity as for example reported in Refs.~\onlinecite{PhysRevLett.123.047401,PhysRevLett.118.215301}.

\section*{acknowledgments}
This work was supported by the Deutsche Forschungsgemeinschaft (DFG) through the collaborative research center TRR142 (project A04, grant No. 231447078) and Heisenberg program (grant No. 270619725) and by the Paderborn Center for Parallel Computing, PC$^2$. X.M. further acknowledges support from the National Natural Science Foundation of China (grant No. 11804064) and the hospitality during his stay at ICFO. Y.V.K. and L.T. acknowledge support from the Severo Ochoa Excellence Programme, Fundacio Cellex, Fundacio Mir-Puig, and CERCA/Generalitat de Catalunya.


\begin{thebibliography}{36}%
\makeatletter
\providecommand \@ifxundefined [1]{%
 \@ifx{#1\undefined}
}%
\providecommand \@ifnum [1]{%
 \ifnum #1\expandafter \@firstoftwo
 \else \expandafter \@secondoftwo
 \fi
}%
\providecommand \@ifx [1]{%
 \ifx #1\expandafter \@firstoftwo
 \else \expandafter \@secondoftwo
 \fi
}%
\providecommand \natexlab [1]{#1}%
\providecommand \enquote  [1]{``#1''}%
\providecommand \bibnamefont  [1]{#1}%
\providecommand \bibfnamefont [1]{#1}%
\providecommand \citenamefont [1]{#1}%
\providecommand \href@noop [0]{\@secondoftwo}%
\providecommand \href [0]{\begingroup \@sanitize@url \@href}%
\providecommand \@href[1]{\@@startlink{#1}\@@href}%
\providecommand \@@href[1]{\endgroup#1\@@endlink}%
\providecommand \@sanitize@url [0]{\catcode `\\12\catcode `\$12\catcode
  `\&12\catcode `\#12\catcode `\^12\catcode `\_12\catcode `\%12\relax}%
\providecommand \@@startlink[1]{}%
\providecommand \@@endlink[0]{}%
\providecommand \url  [0]{\begingroup\@sanitize@url \@url }%
\providecommand \@url [1]{\endgroup\@href {#1}{\urlprefix }}%
\providecommand \urlprefix  [0]{URL }%
\providecommand \Eprint [0]{\href }%
\providecommand \doibase [0]{https://doi.org/}%
\providecommand \selectlanguage [0]{\@gobble}%
\providecommand \bibinfo  [0]{\@secondoftwo}%
\providecommand \bibfield  [0]{\@secondoftwo}%
\providecommand \translation [1]{[#1]}%
\providecommand \BibitemOpen [0]{}%
\providecommand \bibitemStop [0]{}%
\providecommand \bibitemNoStop [0]{.\EOS\space}%
\providecommand \EOS [0]{\spacefactor3000\relax}%
\providecommand \BibitemShut  [1]{\csname bibitem#1\endcsname}%
\let\auto@bib@innerbib\@empty
\bibitem [{\citenamefont {Yarmchuk}\ \emph {et~al.}(1979)\citenamefont
  {Yarmchuk}, \citenamefont {Gordon},\ and\ \citenamefont
  {Packard}}]{yarmchuk1979observation}%
  \BibitemOpen
  \bibfield  {author} {\bibinfo {author} {\bibfnamefont {E.~J.}\ \bibnamefont
  {Yarmchuk}}, \bibinfo {author} {\bibfnamefont {M.~J.~V.}\ \bibnamefont
  {Gordon}},\ and\ \bibinfo {author} {\bibfnamefont {R.~E.}\ \bibnamefont
  {Packard}},\ }\bibfield  {title} {\bibinfo {title} {Observation of stationary
  vortex arrays in rotating superfluid helium},\ }\href
  {https://doi.org/10.1103/PhysRevLett.43.214} {\bibfield  {journal} {\bibinfo
  {journal} {Phys. Rev. Lett.}\ }\textbf {\bibinfo {volume} {43}},\ \bibinfo
  {pages} {214} (\bibinfo {year} {1979})}\BibitemShut {NoStop}%
\bibitem [{\citenamefont {Blatter}\ \emph {et~al.}(1994)\citenamefont
  {Blatter}, \citenamefont {Feigel'man}, \citenamefont {Geshkenbein},
  \citenamefont {Larkin},\ and\ \citenamefont {Vinokur}}]{blatter1994vortices}%
  \BibitemOpen
  \bibfield  {author} {\bibinfo {author} {\bibfnamefont {G.}~\bibnamefont
  {Blatter}}, \bibinfo {author} {\bibfnamefont {M.~V.}\ \bibnamefont
  {Feigel'man}}, \bibinfo {author} {\bibfnamefont {V.~B.}\ \bibnamefont
  {Geshkenbein}}, \bibinfo {author} {\bibfnamefont {A.~I.}\ \bibnamefont
  {Larkin}},\ and\ \bibinfo {author} {\bibfnamefont {V.~M.}\ \bibnamefont
  {Vinokur}},\ }\bibfield  {title} {\bibinfo {title} {Vortices in
  high-temperature superconductors},\ }\href
  {https://doi.org/10.1103/RevModPhys.66.1125} {\bibfield  {journal} {\bibinfo
  {journal} {Rev. Mod. Phys.}\ }\textbf {\bibinfo {volume} {66}},\ \bibinfo
  {pages} {1125} (\bibinfo {year} {1994})}\BibitemShut {NoStop}%
\bibitem [{\citenamefont {Matthews}\ \emph {et~al.}(1999)\citenamefont
  {Matthews}, \citenamefont {Anderson}, \citenamefont {Haljan}, \citenamefont
  {Hall}, \citenamefont {Wieman},\ and\ \citenamefont
  {Cornell}}]{matthews1999vortices}%
  \BibitemOpen
  \bibfield  {author} {\bibinfo {author} {\bibfnamefont {M.~R.}\ \bibnamefont
  {Matthews}}, \bibinfo {author} {\bibfnamefont {B.~P.}\ \bibnamefont
  {Anderson}}, \bibinfo {author} {\bibfnamefont {P.~C.}\ \bibnamefont
  {Haljan}}, \bibinfo {author} {\bibfnamefont {D.~S.}\ \bibnamefont {Hall}},
  \bibinfo {author} {\bibfnamefont {C.~E.}\ \bibnamefont {Wieman}},\ and\
  \bibinfo {author} {\bibfnamefont {E.~A.}\ \bibnamefont {Cornell}},\
  }\bibfield  {title} {\bibinfo {title} {Vortices in a bose-einstein
  condensate},\ }\href {https://doi.org/10.1103/PhysRevLett.83.2498} {\bibfield
   {journal} {\bibinfo  {journal} {Phys. Rev. Lett.}\ }\textbf {\bibinfo
  {volume} {83}},\ \bibinfo {pages} {2498} (\bibinfo {year}
  {1999})}\BibitemShut {NoStop}%
\bibitem [{\citenamefont {Madison}\ \emph {et~al.}(2000)\citenamefont
  {Madison}, \citenamefont {Chevy}, \citenamefont {Wohlleben},\ and\
  \citenamefont {Dalibard}}]{madison2000vortex}%
  \BibitemOpen
  \bibfield  {author} {\bibinfo {author} {\bibfnamefont {K.~W.}\ \bibnamefont
  {Madison}}, \bibinfo {author} {\bibfnamefont {F.}~\bibnamefont {Chevy}},
  \bibinfo {author} {\bibfnamefont {W.}~\bibnamefont {Wohlleben}},\ and\
  \bibinfo {author} {\bibfnamefont {J.}~\bibnamefont {Dalibard}},\ }\bibfield
  {title} {\bibinfo {title} {Vortex formation in a stirred bose-einstein
  condensate},\ }\href {https://doi.org/10.1103/PhysRevLett.84.806} {\bibfield
  {journal} {\bibinfo  {journal} {Phys. Rev. Lett.}\ }\textbf {\bibinfo
  {volume} {84}},\ \bibinfo {pages} {806} (\bibinfo {year} {2000})}\BibitemShut
  {NoStop}%
\bibitem [{\citenamefont {Fetter}\ and\ \citenamefont
  {Svidzinsky}(2001)}]{fetter2001vortices}%
  \BibitemOpen
  \bibfield  {author} {\bibinfo {author} {\bibfnamefont {A.~L.}\ \bibnamefont
  {Fetter}}\ and\ \bibinfo {author} {\bibfnamefont {A.~A.}\ \bibnamefont
  {Svidzinsky}},\ }\bibfield  {title} {\bibinfo {title} {Vortices in a trapped
  dilute bose-einstein condensate},\ }\href@noop {} {\bibfield  {journal}
  {\bibinfo  {journal} {J. of Phys. Condens. Matter}\ }\textbf {\bibinfo
  {volume} {13}},\ \bibinfo {pages} {R135} (\bibinfo {year}
  {2001})}\BibitemShut {NoStop}%
\bibitem [{\citenamefont {Fetter}(2009)}]{fetter2009rotating}%
  \BibitemOpen
  \bibfield  {author} {\bibinfo {author} {\bibfnamefont {A.~L.}\ \bibnamefont
  {Fetter}},\ }\bibfield  {title} {\bibinfo {title} {Rotating trapped
  bose-einstein condensates},\ }\href
  {https://doi.org/10.1103/RevModPhys.81.647} {\bibfield  {journal} {\bibinfo
  {journal} {Rev. Mod. Phys.}\ }\textbf {\bibinfo {volume} {81}},\ \bibinfo
  {pages} {647} (\bibinfo {year} {2009})}\BibitemShut {NoStop}%
\bibitem [{\citenamefont {Swartzlander}\ and\ \citenamefont
  {Law}(1992)}]{swartzlander1992optical}%
  \BibitemOpen
  \bibfield  {author} {\bibinfo {author} {\bibfnamefont {G.~A.}\ \bibnamefont
  {Swartzlander}}\ and\ \bibinfo {author} {\bibfnamefont {C.~T.}\ \bibnamefont
  {Law}},\ }\bibfield  {title} {\bibinfo {title} {Optical vortex solitons
  observed in kerr nonlinear media},\ }\href
  {https://doi.org/10.1103/PhysRevLett.69.2503} {\bibfield  {journal} {\bibinfo
   {journal} {Phys. Rev. Lett.}\ }\textbf {\bibinfo {volume} {69}},\ \bibinfo
  {pages} {2503} (\bibinfo {year} {1992})}\BibitemShut {NoStop}%
\bibitem [{\citenamefont {Desyatnikov}\ \emph {et~al.}(2005)\citenamefont
  {Desyatnikov}, \citenamefont {Kivshar},\ and\ \citenamefont
  {Torner}}]{desyatnikov2005optical}%
  \BibitemOpen
  \bibfield  {author} {\bibinfo {author} {\bibfnamefont {A.~S.}\ \bibnamefont
  {Desyatnikov}}, \bibinfo {author} {\bibfnamefont {Y.~S.}\ \bibnamefont
  {Kivshar}},\ and\ \bibinfo {author} {\bibfnamefont {L.}~\bibnamefont
  {Torner}},\ }\bibfield  {title} {\bibinfo {title} {Optical vortices and
  vortex solitons},\ }\href@noop {} {\bibfield  {journal} {\bibinfo  {journal}
  {Prog. Opt.}\ }\textbf {\bibinfo {volume} {47}},\ \bibinfo {pages} {291}
  (\bibinfo {year} {2005})}\BibitemShut {NoStop}%
\bibitem [{\citenamefont {Lagoudakis}\ \emph {et~al.}(2008)\citenamefont
  {Lagoudakis}, \citenamefont {Wouters}, \citenamefont {Richard}, \citenamefont
  {Baas}, \citenamefont {Carusotto}, \citenamefont {Andr{\'e}}, \citenamefont
  {Dang},\ and\ \citenamefont {Deveaud-Pl{\'e}dran}}]{lagoudakis2008quantized}%
  \BibitemOpen
  \bibfield  {author} {\bibinfo {author} {\bibfnamefont {K.~G.}\ \bibnamefont
  {Lagoudakis}}, \bibinfo {author} {\bibfnamefont {M.}~\bibnamefont {Wouters}},
  \bibinfo {author} {\bibfnamefont {M.}~\bibnamefont {Richard}}, \bibinfo
  {author} {\bibfnamefont {A.}~\bibnamefont {Baas}}, \bibinfo {author}
  {\bibfnamefont {I.}~\bibnamefont {Carusotto}}, \bibinfo {author}
  {\bibfnamefont {R.}~\bibnamefont {Andr{\'e}}}, \bibinfo {author}
  {\bibfnamefont {L.~S.}\ \bibnamefont {Dang}},\ and\ \bibinfo {author}
  {\bibfnamefont {B.}~\bibnamefont {Deveaud-Pl{\'e}dran}},\ }\bibfield  {title}
  {\bibinfo {title} {Quantized vortices in an exciton--polariton condensate},\
  }\href@noop {} {\bibfield  {journal} {\bibinfo  {journal} {Nat. Phys.}\
  }\textbf {\bibinfo {volume} {4}},\ \bibinfo {pages} {706} (\bibinfo {year}
  {2008})}\BibitemShut {NoStop}%
\bibitem [{\citenamefont {Sanvitto}\ \emph {et~al.}(2010)\citenamefont
  {Sanvitto}, \citenamefont {Marchetti}, \citenamefont {Szyma{\'n}ska},
  \citenamefont {Tosi}, \citenamefont {Baudisch}, \citenamefont {Laussy},
  \citenamefont {Krizhanovskii}, \citenamefont {Skolnick}, \citenamefont
  {Marrucci}, \citenamefont {Lema{\^i}tre}, \citenamefont {Bloch},
  \citenamefont {Tejedor},\ and\ \citenamefont
  {Vi{\~n}a}}]{sanvitto2010persistent}%
  \BibitemOpen
  \bibfield  {author} {\bibinfo {author} {\bibfnamefont {D.}~\bibnamefont
  {Sanvitto}}, \bibinfo {author} {\bibfnamefont {F.~M.}\ \bibnamefont
  {Marchetti}}, \bibinfo {author} {\bibfnamefont {M.~H.}\ \bibnamefont
  {Szyma{\'n}ska}}, \bibinfo {author} {\bibfnamefont {G.}~\bibnamefont {Tosi}},
  \bibinfo {author} {\bibfnamefont {M.}~\bibnamefont {Baudisch}}, \bibinfo
  {author} {\bibfnamefont {F.~P.}\ \bibnamefont {Laussy}}, \bibinfo {author}
  {\bibfnamefont {D.~N.}\ \bibnamefont {Krizhanovskii}}, \bibinfo {author}
  {\bibfnamefont {M.~S.}\ \bibnamefont {Skolnick}}, \bibinfo {author}
  {\bibfnamefont {L.}~\bibnamefont {Marrucci}}, \bibinfo {author}
  {\bibfnamefont {A.}~\bibnamefont {Lema{\^i}tre}}, \bibinfo {author}
  {\bibfnamefont {J.}~\bibnamefont {Bloch}}, \bibinfo {author} {\bibfnamefont
  {C.}~\bibnamefont {Tejedor}},\ and\ \bibinfo {author} {\bibfnamefont
  {L.}~\bibnamefont {Vi{\~n}a}},\ }\bibfield  {title} {\bibinfo {title}
  {Persistent currents and quantized vortices in a polariton superfluid},\
  }\href@noop {} {\bibfield  {journal} {\bibinfo  {journal} {Nat. Phys.}\
  }\textbf {\bibinfo {volume} {6}},\ \bibinfo {pages} {527} (\bibinfo {year}
  {2010})}\BibitemShut {NoStop}%
\bibitem [{\citenamefont {Roumpos}\ \emph {et~al.}(2011)\citenamefont
  {Roumpos}, \citenamefont {Fraser}, \citenamefont {L{\"o}ffler}, \citenamefont
  {H{\"o}fling}, \citenamefont {Forchel},\ and\ \citenamefont
  {Yamamoto}}]{roumpos2011single}%
  \BibitemOpen
  \bibfield  {author} {\bibinfo {author} {\bibfnamefont {G.}~\bibnamefont
  {Roumpos}}, \bibinfo {author} {\bibfnamefont {M.~D.}\ \bibnamefont {Fraser}},
  \bibinfo {author} {\bibfnamefont {A.}~\bibnamefont {L{\"o}ffler}}, \bibinfo
  {author} {\bibfnamefont {S.}~\bibnamefont {H{\"o}fling}}, \bibinfo {author}
  {\bibfnamefont {A.}~\bibnamefont {Forchel}},\ and\ \bibinfo {author}
  {\bibfnamefont {Y.}~\bibnamefont {Yamamoto}},\ }\bibfield  {title} {\bibinfo
  {title} {Single vortex--antivortex pair in an exciton-polariton condensate},\
  }\href@noop {} {\bibfield  {journal} {\bibinfo  {journal} {Nat. Phys.}\
  }\textbf {\bibinfo {volume} {7}},\ \bibinfo {pages} {129} (\bibinfo {year}
  {2011})}\BibitemShut {NoStop}%
\bibitem [{\citenamefont {Fedichev}\ and\ \citenamefont
  {Shlyapnikov}(1999)}]{fedichev1999dissipative}%
  \BibitemOpen
  \bibfield  {author} {\bibinfo {author} {\bibfnamefont {P.~O.}\ \bibnamefont
  {Fedichev}}\ and\ \bibinfo {author} {\bibfnamefont {G.~V.}\ \bibnamefont
  {Shlyapnikov}},\ }\bibfield  {title} {\bibinfo {title} {Dissipative dynamics
  of a vortex state in a trapped bose-condensed gas},\ }\href
  {https://doi.org/10.1103/PhysRevA.60.R1779} {\bibfield  {journal} {\bibinfo
  {journal} {Phys. Rev. A}\ }\textbf {\bibinfo {volume} {60}},\ \bibinfo
  {pages} {R1779} (\bibinfo {year} {1999})}\BibitemShut {NoStop}%
\bibitem [{\citenamefont {Jackson}\ \emph {et~al.}(2009)\citenamefont
  {Jackson}, \citenamefont {Proukakis}, \citenamefont {Barenghi},\ and\
  \citenamefont {Zaremba}}]{jackson2009finite}%
  \BibitemOpen
  \bibfield  {author} {\bibinfo {author} {\bibfnamefont {B.}~\bibnamefont
  {Jackson}}, \bibinfo {author} {\bibfnamefont {N.~P.}\ \bibnamefont
  {Proukakis}}, \bibinfo {author} {\bibfnamefont {C.~F.}\ \bibnamefont
  {Barenghi}},\ and\ \bibinfo {author} {\bibfnamefont {E.}~\bibnamefont
  {Zaremba}},\ }\bibfield  {title} {\bibinfo {title} {Finite-temperature vortex
  dynamics in bose-einstein condensates},\ }\href
  {https://doi.org/10.1103/PhysRevA.79.053615} {\bibfield  {journal} {\bibinfo
  {journal} {Phys. Rev. A}\ }\textbf {\bibinfo {volume} {79}},\ \bibinfo
  {pages} {053615} (\bibinfo {year} {2009})}\BibitemShut {NoStop}%
\bibitem [{\citenamefont {Rooney}\ \emph {et~al.}(2010)\citenamefont {Rooney},
  \citenamefont {Bradley},\ and\ \citenamefont {Blakie}}]{rooney2010decay}%
  \BibitemOpen
  \bibfield  {author} {\bibinfo {author} {\bibfnamefont {S.~J.}\ \bibnamefont
  {Rooney}}, \bibinfo {author} {\bibfnamefont {A.~S.}\ \bibnamefont
  {Bradley}},\ and\ \bibinfo {author} {\bibfnamefont {P.~B.}\ \bibnamefont
  {Blakie}},\ }\bibfield  {title} {\bibinfo {title} {Decay of a quantum vortex:
  Test of nonequilibrium theories for warm bose-einstein condensates},\ }\href
  {https://doi.org/10.1103/PhysRevA.81.023630} {\bibfield  {journal} {\bibinfo
  {journal} {Phys. Rev. A}\ }\textbf {\bibinfo {volume} {81}},\ \bibinfo
  {pages} {023630} (\bibinfo {year} {2010})}\BibitemShut {NoStop}%
\bibitem [{\citenamefont {Deng}\ \emph {et~al.}(2002)\citenamefont {Deng},
  \citenamefont {Weihs}, \citenamefont {Santori}, \citenamefont {Bloch},\ and\
  \citenamefont {Yamamoto}}]{deng2002condensation}%
  \BibitemOpen
  \bibfield  {author} {\bibinfo {author} {\bibfnamefont {H.}~\bibnamefont
  {Deng}}, \bibinfo {author} {\bibfnamefont {G.}~\bibnamefont {Weihs}},
  \bibinfo {author} {\bibfnamefont {C.}~\bibnamefont {Santori}}, \bibinfo
  {author} {\bibfnamefont {J.}~\bibnamefont {Bloch}},\ and\ \bibinfo {author}
  {\bibfnamefont {Y.}~\bibnamefont {Yamamoto}},\ }\bibfield  {title} {\bibinfo
  {title} {Condensation of semiconductor microcavity exciton polaritons},\
  }\href@noop {} {\bibfield  {journal} {\bibinfo  {journal} {Science}\ }\textbf
  {\bibinfo {volume} {298}},\ \bibinfo {pages} {199} (\bibinfo {year}
  {2002})}\BibitemShut {NoStop}%
\bibitem [{\citenamefont {Kasprzak}\ \emph {et~al.}(2006)\citenamefont
  {Kasprzak}, \citenamefont {Richard}, \citenamefont {Kundermann},
  \citenamefont {Baas}, \citenamefont {Jeambrun}, \citenamefont {Keeling},
  \citenamefont {Marchetti}, \citenamefont {Szyma{\'n}ska}, \citenamefont
  {Andr{\'e}}, \citenamefont {Staehli}, \citenamefont {Savona}, \citenamefont
  {Littlewood}, \citenamefont {Deveaud},\ and\ \citenamefont
  {Dang}}]{kasprzak2006bose}%
  \BibitemOpen
  \bibfield  {author} {\bibinfo {author} {\bibfnamefont {J.}~\bibnamefont
  {Kasprzak}}, \bibinfo {author} {\bibfnamefont {M.}~\bibnamefont {Richard}},
  \bibinfo {author} {\bibfnamefont {S.}~\bibnamefont {Kundermann}}, \bibinfo
  {author} {\bibfnamefont {A.}~\bibnamefont {Baas}}, \bibinfo {author}
  {\bibfnamefont {P.}~\bibnamefont {Jeambrun}}, \bibinfo {author}
  {\bibfnamefont {J.~M.~J.}\ \bibnamefont {Keeling}}, \bibinfo {author}
  {\bibfnamefont {F.~M.}\ \bibnamefont {Marchetti}}, \bibinfo {author}
  {\bibfnamefont {M.~H.}\ \bibnamefont {Szyma{\'n}ska}}, \bibinfo {author}
  {\bibfnamefont {R.}~\bibnamefont {Andr{\'e}}}, \bibinfo {author}
  {\bibfnamefont {J.~L.}\ \bibnamefont {Staehli}}, \bibinfo {author}
  {\bibfnamefont {V.}~\bibnamefont {Savona}}, \bibinfo {author} {\bibfnamefont
  {P.~B.}\ \bibnamefont {Littlewood}}, \bibinfo {author} {\bibfnamefont
  {B.}~\bibnamefont {Deveaud}},\ and\ \bibinfo {author} {\bibfnamefont {L.~S.}\
  \bibnamefont {Dang}},\ }\bibfield  {title} {\bibinfo {title} {Bose--einstein
  condensation of exciton polaritons},\ }\href@noop {} {\bibfield  {journal}
  {\bibinfo  {journal} {Nature}\ }\textbf {\bibinfo {volume} {443}},\ \bibinfo
  {pages} {409} (\bibinfo {year} {2006})}\BibitemShut {NoStop}%
\bibitem [{\citenamefont {Dominici}\ \emph {et~al.}(2015)\citenamefont
  {Dominici}, \citenamefont {Dagvadorj}, \citenamefont {Fellows}, \citenamefont
  {Ballarini}, \citenamefont {De~Giorgi}, \citenamefont {Marchetti},
  \citenamefont {Piccirillo}, \citenamefont {Marrucci}, \citenamefont
  {Bramati}, \citenamefont {Gigli}, \citenamefont {Szyma{\'n}ska},\ and\
  \citenamefont {Sanvitto}}]{dominici2015vortex}%
  \BibitemOpen
  \bibfield  {author} {\bibinfo {author} {\bibfnamefont {L.}~\bibnamefont
  {Dominici}}, \bibinfo {author} {\bibfnamefont {G.}~\bibnamefont {Dagvadorj}},
  \bibinfo {author} {\bibfnamefont {J.~M.}\ \bibnamefont {Fellows}}, \bibinfo
  {author} {\bibfnamefont {D.}~\bibnamefont {Ballarini}}, \bibinfo {author}
  {\bibfnamefont {M.}~\bibnamefont {De~Giorgi}}, \bibinfo {author}
  {\bibfnamefont {F.~M.}\ \bibnamefont {Marchetti}}, \bibinfo {author}
  {\bibfnamefont {B.}~\bibnamefont {Piccirillo}}, \bibinfo {author}
  {\bibfnamefont {L.}~\bibnamefont {Marrucci}}, \bibinfo {author}
  {\bibfnamefont {A.}~\bibnamefont {Bramati}}, \bibinfo {author} {\bibfnamefont
  {G.}~\bibnamefont {Gigli}}, \bibinfo {author} {\bibfnamefont {M.~H.}\
  \bibnamefont {Szyma{\'n}ska}},\ and\ \bibinfo {author} {\bibfnamefont
  {D.}~\bibnamefont {Sanvitto}},\ }\bibfield  {title} {\bibinfo {title} {Vortex
  and half-vortex dynamics in a nonlinear spinor quantum fluid},\ }\href
  {https://doi.org/10.1126/sciadv.1500807} {\bibfield  {journal} {\bibinfo
  {journal} {Sci. Adv.}\ }\textbf {\bibinfo {volume} {1}},\ \bibinfo {pages}
  {e1500807} (\bibinfo {year} {2015})}\BibitemShut {NoStop}%
\bibitem [{\citenamefont {Kartashov}\ and\ \citenamefont
  {Zezyulin}(2019)}]{zezyulin2019vortex}%
  \BibitemOpen
  \bibfield  {author} {\bibinfo {author} {\bibfnamefont {Y.~V.}\ \bibnamefont
  {Kartashov}}\ and\ \bibinfo {author} {\bibfnamefont {D.~A.}\ \bibnamefont
  {Zezyulin}},\ }\bibfield  {title} {\bibinfo {title} {Rotating patterns in
  polariton condensates in ring-shaped potentials under a bichromatic pump},\
  }\href@noop {} {\bibfield  {journal} {\bibinfo  {journal} {Opt. Lett.}\
  }\textbf {\bibinfo {volume} {44}},\ \bibinfo {pages} {4805} (\bibinfo {year}
  {2019})}\BibitemShut {NoStop}%
\bibitem [{\citenamefont {Dominici}\ \emph {et~al.}(2018)\citenamefont
  {Dominici}, \citenamefont {Colas}, \citenamefont {Gianfrate}, \citenamefont
  {Rahmani}, \citenamefont {Munoz}, \citenamefont {Ballarini}, \citenamefont
  {De~Giorgi}, \citenamefont {Gigli}, \citenamefont {Laussy},\ and\
  \citenamefont {Sanvitto}}]{dominici2018vortex}%
  \BibitemOpen
  \bibfield  {author} {\bibinfo {author} {\bibfnamefont {L.}~\bibnamefont
  {Dominici}}, \bibinfo {author} {\bibfnamefont {D.}~\bibnamefont {Colas}},
  \bibinfo {author} {\bibfnamefont {A.}~\bibnamefont {Gianfrate}}, \bibinfo
  {author} {\bibfnamefont {A.}~\bibnamefont {Rahmani}}, \bibinfo {author}
  {\bibfnamefont {C.~S.}\ \bibnamefont {Munoz}}, \bibinfo {author}
  {\bibfnamefont {D.}~\bibnamefont {Ballarini}}, \bibinfo {author}
  {\bibfnamefont {M.}~\bibnamefont {De~Giorgi}}, \bibinfo {author}
  {\bibfnamefont {G.}~\bibnamefont {Gigli}}, \bibinfo {author} {\bibfnamefont
  {F.~P.}\ \bibnamefont {Laussy}},\ and\ \bibinfo {author} {\bibfnamefont
  {D.}~\bibnamefont {Sanvitto}},\ }\bibfield  {title} {\bibinfo {title}
  {Ultrafast topology shaping by a rabi-oscillating vortex},\ }\href@noop {}
  {\bibfield  {journal} {\bibinfo  {journal} {arXiv preprint arXiv:1801.02580}\
  } (\bibinfo {year} {2018})}\BibitemShut {NoStop}%
\bibitem [{\citenamefont {Dall}\ \emph {et~al.}(2014)\citenamefont {Dall},
  \citenamefont {Fraser}, \citenamefont {Desyatnikov}, \citenamefont {Li},
  \citenamefont {Brodbeck}, \citenamefont {Kamp}, \citenamefont {Schneider},
  \citenamefont {H\"ofling},\ and\ \citenamefont
  {Ostrovskaya}}]{dall2014creation}%
  \BibitemOpen
  \bibfield  {author} {\bibinfo {author} {\bibfnamefont {R.}~\bibnamefont
  {Dall}}, \bibinfo {author} {\bibfnamefont {M.~D.}\ \bibnamefont {Fraser}},
  \bibinfo {author} {\bibfnamefont {A.~S.}\ \bibnamefont {Desyatnikov}},
  \bibinfo {author} {\bibfnamefont {G.}~\bibnamefont {Li}}, \bibinfo {author}
  {\bibfnamefont {S.}~\bibnamefont {Brodbeck}}, \bibinfo {author}
  {\bibfnamefont {M.}~\bibnamefont {Kamp}}, \bibinfo {author} {\bibfnamefont
  {C.}~\bibnamefont {Schneider}}, \bibinfo {author} {\bibfnamefont
  {S.}~\bibnamefont {H\"ofling}},\ and\ \bibinfo {author} {\bibfnamefont
  {E.~A.}\ \bibnamefont {Ostrovskaya}},\ }\bibfield  {title} {\bibinfo {title}
  {Creation of orbital angular momentum states with chiral polaritonic
  lenses},\ }\href {https://doi.org/10.1103/PhysRevLett.113.200404} {\bibfield
  {journal} {\bibinfo  {journal} {Phys. Rev. Lett.}\ }\textbf {\bibinfo
  {volume} {113}},\ \bibinfo {pages} {200404} (\bibinfo {year}
  {2014})}\BibitemShut {NoStop}%
\bibitem [{\citenamefont {Ma}\ \emph {et~al.}(2020)\citenamefont {Ma},
  \citenamefont {Berger}, \citenamefont {A{\ss}mann}, \citenamefont {Driben},
  \citenamefont {Meier}, \citenamefont {Schneider}, \citenamefont
  {H{\"o}fling},\ and\ \citenamefont {Schumacher}}]{ma2020realization}%
  \BibitemOpen
  \bibfield  {author} {\bibinfo {author} {\bibfnamefont {X.}~\bibnamefont
  {Ma}}, \bibinfo {author} {\bibfnamefont {B.}~\bibnamefont {Berger}}, \bibinfo
  {author} {\bibfnamefont {M.}~\bibnamefont {A{\ss}mann}}, \bibinfo {author}
  {\bibfnamefont {R.}~\bibnamefont {Driben}}, \bibinfo {author} {\bibfnamefont
  {T.}~\bibnamefont {Meier}}, \bibinfo {author} {\bibfnamefont
  {C.}~\bibnamefont {Schneider}}, \bibinfo {author} {\bibfnamefont
  {S.}~\bibnamefont {H{\"o}fling}},\ and\ \bibinfo {author} {\bibfnamefont
  {S.}~\bibnamefont {Schumacher}},\ }\bibfield  {title} {\bibinfo {title}
  {Realization of all-optical vortex switching in exciton-polariton
  condensates},\ }\href@noop {} {\bibfield  {journal} {\bibinfo  {journal}
  {Nat. Commun.}\ }\textbf {\bibinfo {volume} {11}},\ \bibinfo {pages} {897}
  (\bibinfo {year} {2020})}\BibitemShut {NoStop}%
\bibitem [{\citenamefont {Liew}\ \emph {et~al.}(2015)\citenamefont {Liew},
  \citenamefont {Egorov}, \citenamefont {Matuszewski}, \citenamefont
  {Kyriienko}, \citenamefont {Ma},\ and\ \citenamefont
  {Ostrovskaya}}]{liew2015instability}%
  \BibitemOpen
  \bibfield  {author} {\bibinfo {author} {\bibfnamefont {T.~C.~H.}\
  \bibnamefont {Liew}}, \bibinfo {author} {\bibfnamefont {O.~A.}\ \bibnamefont
  {Egorov}}, \bibinfo {author} {\bibfnamefont {M.}~\bibnamefont {Matuszewski}},
  \bibinfo {author} {\bibfnamefont {O.}~\bibnamefont {Kyriienko}}, \bibinfo
  {author} {\bibfnamefont {X.}~\bibnamefont {Ma}},\ and\ \bibinfo {author}
  {\bibfnamefont {E.~A.}\ \bibnamefont {Ostrovskaya}},\ }\bibfield  {title}
  {\bibinfo {title} {Instability-induced formation and nonequilibrium dynamics
  of phase defects in polariton condensates},\ }\href
  {https://doi.org/10.1103/PhysRevB.91.085413} {\bibfield  {journal} {\bibinfo
  {journal} {Phys. Rev. B}\ }\textbf {\bibinfo {volume} {91}},\ \bibinfo
  {pages} {085413} (\bibinfo {year} {2015})}\BibitemShut {NoStop}%
\bibitem [{\citenamefont {Warszawski}\ \emph {et~al.}(2012)\citenamefont
  {Warszawski}, \citenamefont {Melatos},\ and\ \citenamefont
  {Berloff}}]{warszawski2012unpinning}%
  \BibitemOpen
  \bibfield  {author} {\bibinfo {author} {\bibfnamefont {L.}~\bibnamefont
  {Warszawski}}, \bibinfo {author} {\bibfnamefont {A.}~\bibnamefont
  {Melatos}},\ and\ \bibinfo {author} {\bibfnamefont {N.~G.}\ \bibnamefont
  {Berloff}},\ }\bibfield  {title} {\bibinfo {title} {Unpinning triggers for
  superfluid vortex avalanches},\ }\href
  {https://doi.org/10.1103/PhysRevB.85.104503} {\bibfield  {journal} {\bibinfo
  {journal} {Phys. Rev. B}\ }\textbf {\bibinfo {volume} {85}},\ \bibinfo
  {pages} {104503} (\bibinfo {year} {2012})}\BibitemShut {NoStop}%
\bibitem [{\citenamefont {Ostrovskaya}\ \emph {et~al.}(2012)\citenamefont
  {Ostrovskaya}, \citenamefont {Abdullaev}, \citenamefont {Desyatnikov},
  \citenamefont {Fraser},\ and\ \citenamefont
  {Kivshar}}]{ostrovskaya2012dissipative}%
  \BibitemOpen
  \bibfield  {author} {\bibinfo {author} {\bibfnamefont {E.~A.}\ \bibnamefont
  {Ostrovskaya}}, \bibinfo {author} {\bibfnamefont {J.}~\bibnamefont
  {Abdullaev}}, \bibinfo {author} {\bibfnamefont {A.~S.}\ \bibnamefont
  {Desyatnikov}}, \bibinfo {author} {\bibfnamefont {M.~D.}\ \bibnamefont
  {Fraser}},\ and\ \bibinfo {author} {\bibfnamefont {Y.~S.}\ \bibnamefont
  {Kivshar}},\ }\bibfield  {title} {\bibinfo {title} {Dissipative solitons and
  vortices in polariton bose-einstein condensates},\ }\href
  {https://doi.org/10.1103/PhysRevA.86.013636} {\bibfield  {journal} {\bibinfo
  {journal} {Phys. Rev. A}\ }\textbf {\bibinfo {volume} {86}},\ \bibinfo
  {pages} {013636} (\bibinfo {year} {2012})}\BibitemShut {NoStop}%
\bibitem [{\citenamefont {Borgh}\ \emph {et~al.}(2012)\citenamefont {Borgh},
  \citenamefont {Franchetti}, \citenamefont {Keeling},\ and\ \citenamefont
  {Berloff}}]{borgh2012robustness}%
  \BibitemOpen
  \bibfield  {author} {\bibinfo {author} {\bibfnamefont {M.~O.}\ \bibnamefont
  {Borgh}}, \bibinfo {author} {\bibfnamefont {G.}~\bibnamefont {Franchetti}},
  \bibinfo {author} {\bibfnamefont {J.}~\bibnamefont {Keeling}},\ and\ \bibinfo
  {author} {\bibfnamefont {N.~G.}\ \bibnamefont {Berloff}},\ }\bibfield
  {title} {\bibinfo {title} {Robustness and observability of rotating vortex
  lattices in an exciton-polariton condensate},\ }\href
  {https://doi.org/10.1103/PhysRevB.86.035307} {\bibfield  {journal} {\bibinfo
  {journal} {Phys. Rev. B}\ }\textbf {\bibinfo {volume} {86}},\ \bibinfo
  {pages} {035307} (\bibinfo {year} {2012})}\BibitemShut {NoStop}%
\bibitem [{\citenamefont {Rosanov}\ \emph {et~al.}(2005)\citenamefont
  {Rosanov}, \citenamefont {Fedorov},\ and\ \citenamefont
  {Shatsev}}]{rosanov2005curvilinear}%
  \BibitemOpen
  \bibfield  {author} {\bibinfo {author} {\bibfnamefont {N.~N.}\ \bibnamefont
  {Rosanov}}, \bibinfo {author} {\bibfnamefont {S.~V.}\ \bibnamefont
  {Fedorov}},\ and\ \bibinfo {author} {\bibfnamefont {A.~N.}\ \bibnamefont
  {Shatsev}},\ }\bibfield  {title} {\bibinfo {title} {Curvilinear motion of
  multivortex laser-soliton complexes with strong and weak coupling},\ }\href
  {https://doi.org/10.1103/PhysRevLett.95.053903} {\bibfield  {journal}
  {\bibinfo  {journal} {Phys. Rev. Lett.}\ }\textbf {\bibinfo {volume} {95}},\
  \bibinfo {pages} {053903} (\bibinfo {year} {2005})}\BibitemShut {NoStop}%
\bibitem [{\citenamefont {Fraser}\ \emph {et~al.}(2009)\citenamefont {Fraser},
  \citenamefont {Roumpos},\ and\ \citenamefont {Yamamoto}}]{fraser2009vortex}%
  \BibitemOpen
  \bibfield  {author} {\bibinfo {author} {\bibfnamefont {M.~D.}\ \bibnamefont
  {Fraser}}, \bibinfo {author} {\bibfnamefont {G.}~\bibnamefont {Roumpos}},\
  and\ \bibinfo {author} {\bibfnamefont {Y.}~\bibnamefont {Yamamoto}},\
  }\bibfield  {title} {\bibinfo {title} {Vortex--antivortex pair dynamics in an
  exciton--polariton condensate},\ }\href@noop {} {\bibfield  {journal}
  {\bibinfo  {journal} {New J. of Phys.}\ }\textbf {\bibinfo {volume} {11}},\
  \bibinfo {pages} {113048} (\bibinfo {year} {2009})}\BibitemShut {NoStop}%
\bibitem [{\citenamefont {Nardin}\ \emph {et~al.}(2011)\citenamefont {Nardin},
  \citenamefont {Grosso}, \citenamefont {L{\'e}ger}, \citenamefont
  {Pi{\c{e}}tka}, \citenamefont {Morier-Genoud},\ and\ \citenamefont
  {Deveaud-Pl{\'e}dran}}]{nardin2011hydrodynamic}%
  \BibitemOpen
  \bibfield  {author} {\bibinfo {author} {\bibfnamefont {G.}~\bibnamefont
  {Nardin}}, \bibinfo {author} {\bibfnamefont {G.}~\bibnamefont {Grosso}},
  \bibinfo {author} {\bibfnamefont {Y.}~\bibnamefont {L{\'e}ger}}, \bibinfo
  {author} {\bibfnamefont {B.}~\bibnamefont {Pi{\c{e}}tka}}, \bibinfo {author}
  {\bibfnamefont {F.}~\bibnamefont {Morier-Genoud}},\ and\ \bibinfo {author}
  {\bibfnamefont {B.}~\bibnamefont {Deveaud-Pl{\'e}dran}},\ }\bibfield  {title}
  {\bibinfo {title} {Hydrodynamic nucleation of quantized vortex pairs in a
  polariton quantum fluid},\ }\href@noop {} {\bibfield  {journal} {\bibinfo
  {journal} {Nat. Phys.}\ }\textbf {\bibinfo {volume} {7}},\ \bibinfo {pages}
  {635} (\bibinfo {year} {2011})}\BibitemShut {NoStop}%
\bibitem [{\citenamefont {Hendrey}\ \emph {et~al.}(2000)\citenamefont
  {Hendrey}, \citenamefont {Nam}, \citenamefont {Guzdar},\ and\ \citenamefont
  {Ott}}]{PhysRevE.62.7627}%
  \BibitemOpen
  \bibfield  {author} {\bibinfo {author} {\bibfnamefont {M.}~\bibnamefont
  {Hendrey}}, \bibinfo {author} {\bibfnamefont {K.}~\bibnamefont {Nam}},
  \bibinfo {author} {\bibfnamefont {P.}~\bibnamefont {Guzdar}},\ and\ \bibinfo
  {author} {\bibfnamefont {E.}~\bibnamefont {Ott}},\ }\bibfield  {title}
  {\bibinfo {title} {Target waves in the complex ginzburg-landau equation},\
  }\href {https://doi.org/10.1103/PhysRevE.62.7627} {\bibfield  {journal}
  {\bibinfo  {journal} {Phys. Rev. E}\ }\textbf {\bibinfo {volume} {62}},\
  \bibinfo {pages} {7627} (\bibinfo {year} {2000})}\BibitemShut {NoStop}%
\bibitem [{\citenamefont {Aranson}\ and\ \citenamefont
  {Kramer}(2002)}]{RevModPhys.74.99}%
  \BibitemOpen
  \bibfield  {author} {\bibinfo {author} {\bibfnamefont {I.~S.}\ \bibnamefont
  {Aranson}}\ and\ \bibinfo {author} {\bibfnamefont {L.}~\bibnamefont
  {Kramer}},\ }\bibfield  {title} {\bibinfo {title} {The world of the complex
  ginzburg-landau equation},\ }\href {https://doi.org/10.1103/RevModPhys.74.99}
  {\bibfield  {journal} {\bibinfo  {journal} {Rev. Mod. Phys.}\ }\textbf
  {\bibinfo {volume} {74}},\ \bibinfo {pages} {99} (\bibinfo {year}
  {2002})}\BibitemShut {NoStop}%
\bibitem [{\citenamefont {Wouters}\ and\ \citenamefont
  {Carusotto}(2007)}]{wouters2007excitations}%
  \BibitemOpen
  \bibfield  {author} {\bibinfo {author} {\bibfnamefont {M.}~\bibnamefont
  {Wouters}}\ and\ \bibinfo {author} {\bibfnamefont {I.}~\bibnamefont
  {Carusotto}},\ }\bibfield  {title} {\bibinfo {title} {Excitations in a
  nonequilibrium bose-einstein condensate of exciton polaritons},\ }\href
  {https://doi.org/10.1103/PhysRevLett.99.140402} {\bibfield  {journal}
  {\bibinfo  {journal} {Phys. Rev. Lett.}\ }\textbf {\bibinfo {volume} {99}},\
  \bibinfo {pages} {140402} (\bibinfo {year} {2007})}\BibitemShut {NoStop}%
\bibitem [{vid()}]{videos}%
  \BibitemOpen
  \href@noop {} {\ }\bibinfo {note} {See Supplementary Material at http:// for
  the videos of the time evolution of multiple spiraling vortice in Fig.
  7(c)-7(e) and the time evolution of a single vortex as pump intensity
  changes.}\BibitemShut {Stop}%
\bibitem [{\citenamefont {Ma}\ \emph {et~al.}(2016)\citenamefont {Ma},
  \citenamefont {Peschel},\ and\ \citenamefont {Egorov}}]{ma2016incoherent}%
  \BibitemOpen
  \bibfield  {author} {\bibinfo {author} {\bibfnamefont {X.}~\bibnamefont
  {Ma}}, \bibinfo {author} {\bibfnamefont {U.}~\bibnamefont {Peschel}},\ and\
  \bibinfo {author} {\bibfnamefont {O.~A.}\ \bibnamefont {Egorov}},\ }\bibfield
   {title} {\bibinfo {title} {Incoherent control of topological charges in
  nonequilibrium polariton condensates},\ }\href
  {https://doi.org/10.1103/PhysRevB.93.035315} {\bibfield  {journal} {\bibinfo
  {journal} {Phys. Rev. B}\ }\textbf {\bibinfo {volume} {93}},\ \bibinfo
  {pages} {035315} (\bibinfo {year} {2016})}\BibitemShut {NoStop}%
\bibitem [{\citenamefont {Kwon}\ \emph {et~al.}(2019)\citenamefont {Kwon},
  \citenamefont {Oh}, \citenamefont {Gong}, \citenamefont {Kim}, \citenamefont
  {Kang}, \citenamefont {Kang}, \citenamefont {Song}, \citenamefont {Choi},\
  and\ \citenamefont {Cho}}]{kwon2019direct}%
  \BibitemOpen
  \bibfield  {author} {\bibinfo {author} {\bibfnamefont {M.-S.}\ \bibnamefont
  {Kwon}}, \bibinfo {author} {\bibfnamefont {B.~Y.}\ \bibnamefont {Oh}},
  \bibinfo {author} {\bibfnamefont {S.-H.}\ \bibnamefont {Gong}}, \bibinfo
  {author} {\bibfnamefont {J.-H.}\ \bibnamefont {Kim}}, \bibinfo {author}
  {\bibfnamefont {H.~K.}\ \bibnamefont {Kang}}, \bibinfo {author}
  {\bibfnamefont {S.}~\bibnamefont {Kang}}, \bibinfo {author} {\bibfnamefont
  {J.~D.}\ \bibnamefont {Song}}, \bibinfo {author} {\bibfnamefont
  {H.}~\bibnamefont {Choi}},\ and\ \bibinfo {author} {\bibfnamefont {Y.-H.}\
  \bibnamefont {Cho}},\ }\bibfield  {title} {\bibinfo {title} {Direct transfer
  of light's orbital angular momentum onto a nonresonantly excited polariton
  superfluid},\ }\href {https://doi.org/10.1103/PhysRevLett.122.045302}
  {\bibfield  {journal} {\bibinfo  {journal} {Phys. Rev. Lett.}\ }\textbf
  {\bibinfo {volume} {122}},\ \bibinfo {pages} {045302} (\bibinfo {year}
  {2019})}\BibitemShut {NoStop}%
\bibitem [{\citenamefont {Ballarini}\ \emph {et~al.}(2019)\citenamefont
  {Ballarini}, \citenamefont {Chestnov}, \citenamefont {Caputo}, \citenamefont
  {De~Giorgi}, \citenamefont {Dominici}, \citenamefont {West}, \citenamefont
  {Pfeiffer}, \citenamefont {Gigli}, \citenamefont {Kavokin},\ and\
  \citenamefont {Sanvitto}}]{PhysRevLett.123.047401}%
  \BibitemOpen
  \bibfield  {author} {\bibinfo {author} {\bibfnamefont {D.}~\bibnamefont
  {Ballarini}}, \bibinfo {author} {\bibfnamefont {I.}~\bibnamefont {Chestnov}},
  \bibinfo {author} {\bibfnamefont {D.}~\bibnamefont {Caputo}}, \bibinfo
  {author} {\bibfnamefont {M.}~\bibnamefont {De~Giorgi}}, \bibinfo {author}
  {\bibfnamefont {L.}~\bibnamefont {Dominici}}, \bibinfo {author}
  {\bibfnamefont {K.}~\bibnamefont {West}}, \bibinfo {author} {\bibfnamefont
  {L.~N.}\ \bibnamefont {Pfeiffer}}, \bibinfo {author} {\bibfnamefont
  {G.}~\bibnamefont {Gigli}}, \bibinfo {author} {\bibfnamefont
  {A.}~\bibnamefont {Kavokin}},\ and\ \bibinfo {author} {\bibfnamefont
  {D.}~\bibnamefont {Sanvitto}},\ }\bibfield  {title} {\bibinfo {title}
  {Self-trapping of exciton-polariton condensates in gaas microcavities},\
  }\href {https://doi.org/10.1103/PhysRevLett.123.047401} {\bibfield  {journal}
  {\bibinfo  {journal} {Phys. Rev. Lett.}\ }\textbf {\bibinfo {volume} {123}},\
  \bibinfo {pages} {047401} (\bibinfo {year} {2019})}\BibitemShut {NoStop}%
\bibitem [{\citenamefont {Ballarini}\ \emph {et~al.}(2017)\citenamefont
  {Ballarini}, \citenamefont {Caputo}, \citenamefont {Mu\~noz}, \citenamefont
  {De~Giorgi}, \citenamefont {Dominici}, \citenamefont
  {Szyma\ifmmode~\acute{n}\else \'{n}\fi{}ska}, \citenamefont {West},
  \citenamefont {Pfeiffer}, \citenamefont {Gigli}, \citenamefont {Laussy},\
  and\ \citenamefont {Sanvitto}}]{PhysRevLett.118.215301}%
  \BibitemOpen
  \bibfield  {author} {\bibinfo {author} {\bibfnamefont {D.}~\bibnamefont
  {Ballarini}}, \bibinfo {author} {\bibfnamefont {D.}~\bibnamefont {Caputo}},
  \bibinfo {author} {\bibfnamefont {C.~S.}\ \bibnamefont {Mu\~noz}}, \bibinfo
  {author} {\bibfnamefont {M.}~\bibnamefont {De~Giorgi}}, \bibinfo {author}
  {\bibfnamefont {L.}~\bibnamefont {Dominici}}, \bibinfo {author}
  {\bibfnamefont {M.~H.}\ \bibnamefont {Szyma\ifmmode~\acute{n}\else
  \'{n}\fi{}ska}}, \bibinfo {author} {\bibfnamefont {K.}~\bibnamefont {West}},
  \bibinfo {author} {\bibfnamefont {L.~N.}\ \bibnamefont {Pfeiffer}}, \bibinfo
  {author} {\bibfnamefont {G.}~\bibnamefont {Gigli}}, \bibinfo {author}
  {\bibfnamefont {F.~P.}\ \bibnamefont {Laussy}},\ and\ \bibinfo {author}
  {\bibfnamefont {D.}~\bibnamefont {Sanvitto}},\ }\bibfield  {title} {\bibinfo
  {title} {Macroscopic two-dimensional polariton condensates},\ }\href
  {https://doi.org/10.1103/PhysRevLett.118.215301} {\bibfield  {journal}
  {\bibinfo  {journal} {Phys. Rev. Lett.}\ }\textbf {\bibinfo {volume} {118}},\
  \bibinfo {pages} {215301} (\bibinfo {year} {2017})}\BibitemShut {NoStop}%
\end{thebibliography}

%

\end{document}